\documentclass[aps,
prd,
preprint,
eqsecnum,
amsmath,
amssymb
]{revtex4}

\usepackage{hyperref}
\usepackage{color}
\usepackage{mathrsfs}
\usepackage{xcolor}
\usepackage{graphicx}

\begin{document}

\title{Electrovacuum solutions in non-local gravity}

\author{Karan Fernandes} \email[email: ]{karan12t@bose.res.in}
\affiliation{S N Bose National Centre for Basic Sciences, Block JD, Sector III, Salt Lake, Kolkata 700106, India}

\author{Arpita Mitra} \email[email: ]{arpita12t@bose.res.in}
\affiliation{S N Bose National Centre for Basic Sciences, Block JD, Sector III, Salt Lake, Kolkata 700106, India}

\begin{abstract}
We consider the coupling of the electromagnetic field to a non-local gravity theory comprising of the Einstein-Hilbert action in addition to a non-local $R\, {\Box}^{-2} R$ term associated with a mass scale $m$. 
We demonstrate that in the case of the minimally coupled electromagnetic field, real corrections about the Reissner-Nordstr\"om background only exist between the inner Cauchy horizon and the event horizon of the black hole. This motivates us to consider the modified coupling of electromagnetism to this theory via the Kaluza ansatz.
The Kaluza reduction introduces non-local terms involving the electromagnetic field to the pure gravitational non-local theory. An iterative approach is provided to perturbatively solve the equations of motion to arbitrary order in $m^2$ about any known solution of General Relativity. We derive the first-order corrections and demonstrate that the higher order corrections are real and perturbative about the external background of a Reissner-Nordstr\"om black hole. We also discuss how the Kaluza reduced action, through the inclusion of non-local electromagnetic fields, could also be relevant in quantum effects on curved backgrounds with horizons. 
\end{abstract}


\maketitle

\section{Introduction}

The observed late-time accelerated expansion of our Universe \cite{Perlmutter:1998np, Riess:1998cb} have inspired numerous modified gravity theories which aim to provide a natural explanation for the cosmological constant. One such class of theories which have been considerably investigated over the years include non-local corrections of either the Einstein field equations or the Einstein-Hilbert action. Non-locality has been used to address several areas in cosmology and black hole physics, including an explanation for dark energy~\cite{ArkaniHamed:2002fu,Barvinsky:2003kg,Dvali:2006su,Dvali:2007kt,Deser:2007jk,Nojiri:2007uq,
Koivisto:2008xfa,Capozziello:2008gu,Deffayet:2009ca,Elizalde:2011su,Foffa:2013vma,Sandstad:2013oja,Woodard:2014iga,Barvinsky:2014lja,Tsamis:2014hra}, dark matter~\cite{Soussa:2003vv,Hehl:2008eu,Barvinsky:2011hd,Arraut:2013qra,Deffayet:2014lba,Woodard:2014wia}, path integrals in quantum gravity~\cite{Wetterich:1997bz, Donoghue:2015nba}, ghost-free higher derivative theories~\cite{Tomboulis:1997gg,Biswas:2011ar,Modesto:2011kw,Modesto:2012ys,Biswas:2013cha,Briscese:2013lna,Modesto:2014lga,Talaganis:2014ida,Tomboulis:2015esa,Modesto:2017sdr} and the near horizon properties of black holes~\cite{Giddings:2009ae,Giddings:2012gc,Giddings:2017mym,Liebling:2017pqs}, among other areas. Non-local terms generally result in effective actions due to integrating out certain fields from a given theory. Many of the models in the references above were in part inspired by the anomaly-induced quantum effective actions resulting from gravitational anomalies and their applications~\cite{Buchbinder:1992rb, Shapiro:2008sf}. Through their involvement of inverse differential operators, non-local terms are also associated with the infrared corrections of a theory. 
However, the construction of a consistent infrared deformation of General Relativity (GR) provides a considerable challenge. The theory should be consistent with current cosmological observations, respect causality 
and be free of ghosts \cite{Deser:2013uya, Maggiore:2016gpx}, at least up to a reasonable UV  cutoff. 
One of the earliest proposed modifications involves the degravitation idea~\cite{ArkaniHamed:2002fu, Dvali:2006su}, where non-local corrections of the Einstein field equations involving the inverse D'Alembertian could filter out the contribution of the vaccuum energy density to the cosmological constant. These equations were later shown to admit a stress tensor which is in general not conserved on curved backgrounds. The first covariant description following the degravitation idea was provided in \cite{Foffa:2013vma},
%
%
\begin{equation}
G_{\mu\nu}-\frac{m^2}{3}(g_{\mu\nu}\Box^{-1}R)^T=8\pi GT_{\mu\nu}
\label{eom.mag1}
\end{equation}
where $m$ is a mass scale taken to be of the order of the Hubble parameter $(H_0)$, $\Box = g^{\mu \nu} \nabla_{\mu} \nabla_{\nu}$ is the D'Alembertian operator and the superscript `$T$' denotes the transverse component. By explicitly considering the transverse component, the non-local term in Eq.~(\ref{eom.mag1}) ensures the covariant conservation of the stress tensor and that the ghosts remain non-radiative. 
While a covariant action from which this equation follows has not yet been derived, another action where the equations agree at the linearized level was introduced in~\cite{Maggiore:2014sia}
\begin{equation}
S_{MM}= \frac{1}{16 \pi G}\int~d^4x \sqrt{-g}\left[R-\frac{m^2}{6}R\frac{1}{\Box^2}R\right]
\label{act.mag1}
\end{equation}
 Eq.~(\ref{eom.mag1}), as well as the equation of motion resulting from Eq.~(\ref{act.mag1}), have been shown to be consistent with solar system tests and agree, as well as the $\Lambda$CDM model, with current cosmological observations~\cite{Nesseris:2014mea,Barreira:2014kra,Dirian:2014xoa,Dirian:2014ara,Dirian:2014bma,Maggiore:2014sia,Dirian:2016puz,Nersisyan:2016hjh, Maggiore:2016gpx,Belgacem:2017cqo}. The spherically symmetric corrections to the Schwarzschild and FRW backgrounds were also derived in \cite{Maggiore:2014sia}. These solutions do not involve non-linear instabilities in the region outside the horizon. This is unlike the scenario in massive gravity theories generically, where non-linearities do creep in below a Vainshtein radius thereby causing a breakdown of the theory well outside of the event horizon. The success of non-local theories such as Eq.~(\ref{act.mag1}) however are not known to include general theories with tensorial non-localities. Theories involving curvature terms other than the Ricci scalar generally do not lead to stable cosmological evolution \cite{Nersisyan:2016jta}.



Non-local field theories can always be treated as local constrained field theories through the introduction of non-dynamical auxilliary fields~\cite{Maggiore:2016gpx}. The non-local terms of the original theory impose constraints on the auxilliary fields which hold for a large class of non-local theories. For instance, let us consider the case of a gravitational action which involves a $f(R)\Box^{-n}R$ term, where $f(R)$ is an arbitrary function of the Ricci scalar and `$n$' is an arbitrary positive integer. The local formulation of this action will then always involve an auxilliary field $U$ satisfying $\Box U = -R$. This field plays a central role in our consideration of the coupling of the electromagnetic field. Thus while our arguments will be made for  Eq.~(\ref{act.mag1}), our analysis will hold for a large class of non-local pure gravitational theories, particularly for all local formulations of non-local theories containing $U$ as defined. 

In the case of local field theories, it is natural to consider the minimal coupling presciption. However, as just discussed, the dynamics of non-local theories requires a local formulation with certain constraints which the gravitational fields must satisfy. One could thus consider the non-local coupling of matter fields such that the constraints include the matter fields as well.
Such non-local terms which involve both gravitational and gauge fields appear regularly in the context of anomaly-induced quantum effective actions. In this paper, we will use the Kaluza ansatz as a prescription to include non-local electromagnetic terms in a given non-local pure gravitational action. We apply the Kaluza ansatz to Eq.~(\ref{act.mag1}) in five-dimensions following the approach considered in \cite{Fernandes:2014bka}. The Kaluza-Klein reduction in the context of other modified gravity theories \cite{Fernandes:2014bka, Bamba:2013fta} results in an effective action involving non-minimal couplings between the electromagnetic field strength and curvature tensors. Non-minimal couplings between the electromagnetic field and curvature terms have been considered previously for their cosmological implications~\cite{Ashoorioon:2004rs,Golovnev:2008cf,Bamba:2008ja,Sadeghi:2009pu,Harko:2014gwa}. They also arise in the one-loop effective action for quantum electrodynamics on curved backgrounds~\cite{Drummond:1979pp}, which have applications in graviton-photon scattering processes~\cite{Bastianelli:2004zp,Bastianelli:2007jv,Bastianelli:2012bz}. The action resulting from the Kaluza ansatz in the present case will lead to terms non-local in both the electromagnetic field strength tensor and the Ricci scalar.

In considering corrections about the Reissner-Nordstr\"om (RN) background, we will demonstrate that the usual minimal coupling of the electromagnetic field is unsatisfactory. Apart from being complex outside the event horizon of the black hole, the corrections do not reduce to those noted for the Schwarzschild background in the limit of vanishing charge. Thus while real perturbative corrections can be derived from Eq.~(\ref{act.mag1}) about vacuum solutions of GR, they do not appear to exist about known electrovacuum solutions of GR in the minimally coupled case. In contrast, the corrections to the RN background resulting from the Kaluza reduced action are well defined in the region beyond the event horizon and reduce to the known corrections for the Schwarzschild background in the limit of vanishing charge. 

The organization of our paper is as follows. In Sec.~\ref{sec.mm}, we review the basic properties of the action introduced in \cite{Maggiore:2014sia} and the correction derived about the Schwarzschild background. We then demonstrate that such real corrections do not exist outside the event horizon of a RN black hole when the electromagnetic field is minimally coupled to the theory. In Sec.~\ref{sec.kk}, we apply the Kaluza ansatz to the five-dimensional gravitational action to derive an effective action which is non-local in both electromagnetic and gravitational fields. In Sec.~\ref{sec.eom}, the equations of motion of this action are derived and compared with those for the original non-local theory with a minimally coupled electromagnetic field. Sec.~\ref{sec.sol} describes an iterative procedure which can be used to solve the equations of motion resulting from the Kaluza reduced non-local action. Here we derive the first order corrections about the RN background and consider the form of the higher order corrections. From this we argue that perturbative corrections exist to all orders about the RN background. We then conclude with a discussion of our results and future directions in Sec.~\ref{sec.con}.

\section{The Non-Local gravitational action} \label{sec.mm}
In this section, we will review the non-local action, its equations of motion and the derivation of the corrections about the Schwarzschild background as provided in~\cite{Kehagias:2014sda,Maggiore:2014sia}. We will denote the gravitational action introduced in \cite{Maggiore:2014sia} by $S_{MM}$
\begin{equation}
S_{MM} =  \frac{1}{16 \pi G}\int d^4x \sqrt{-g} \left[R - \mu R \frac{1}{\Box^2}R \right] \, ,
\label{mag.act1}
\end{equation}
where $G$ is the Newton constant, $g$ is the determinant of the metric $g_{\mu \nu}$, $\mu = \frac{m^2}{6}$ is the mass term associated with the additional non-local contribution in the action and $\Box$ is the D'Alembertian operator. For the rest of the paper, we will set $G=1$. We can always include a given local matter action $S_M$ to Eq.~(\ref{mag.act1}) to define
\begin{align}
S_{NL} &= S_{MM} + S_M \, .
\label{mag.act}
\end{align}
We can further define two auxilliary, non-dynamical fields $U$ and $S$ which satisfy
\begin{equation}
\Box~ U = - R \, , \quad \Box~ S = - U \, ,
\label{mag.aux}
\end{equation}
These fields can be included in Eq.~(\ref{mag.act}) with the help of Lagrange multipliers $\xi_1$ and $\xi_2$ to provide the following local formulation
\begin{equation}
S_{NL} = \frac{1}{16 \pi }\int d^4x \sqrt{-g} \left[R - \mu R S -\xi_1(\Box U + R) - \xi_2(\Box S + U)\right] + S_M \, .
\label{mag.act2}
\end{equation}
The equations of motion for $U$ and $S$ are given by
\begin{align}
\Box ~\xi_1 + \xi_2 &= 0 \, , \notag\\
\Box ~\xi_2 + \mu R &= 0 \, ,
\label{eom.mult}
\end{align}
respectively, from which we can identify $\xi_2 = \mu U$ and $\xi_1 = \mu S$ on comparing with Eq.~(\ref{mag.aux}). With these expressions for $\xi_1$ and $\xi_2$, we have the following equation of motion for the metric
\begin{equation}
G_{\mu \nu}(1 - 2\mu S) -  8 \pi T_{\mu \nu} = \mu K_{\mu \nu} \, ,
\label{eom.met}
\end{equation}
where the stress energy tensor $T_{\mu \nu}$ is defined in the usual way
\begin{equation}
T_{\mu \nu} := - \frac{2}{\sqrt{- g}} \frac{\delta S_M}{\delta g^{\mu \nu}} \,  
\label{eom.set}
\end{equation}
and $K_{\mu \nu}$ is given by
\begin{equation}
K_{\mu \nu} = g_{\mu \nu} \left(2 \Box S + \nabla_{\alpha}U\nabla^{\alpha}S - \frac{1}{2}U^2 \right)  - 2 \nabla_{\mu} \nabla_{\nu}S - \left(\nabla_{\mu}U \nabla_{\nu}S + \nabla_{\nu}U \nabla_{\mu}S\right) \,.
\label{eom.corr}
\end{equation}
It follows from Eq.~(\ref{eom.corr}) that $\nabla_{\mu}K^{\mu \nu} = 0$. Since the matter theory is minimally coupled to the background, its equations of motion are unaffected in the present case. Thus the modified theory satisfies the usual conservation equations. The trace of the field equations on the other hand is modified. Denoting the trace of the stress-energy tensor as $T$, the trace of Eq.~(\ref{eom.met}) is given by 
\begin{equation}
R(1- 2\mu S) + 8\pi T = -\mu \left(6 \Box S - 2 U^2 + 2 \nabla_{\alpha}U \nabla^{\alpha}S \right) \, ,
\label{eom.tr}
\end{equation}
\subsection{Corrections about the Schwarzschild background}
The field equations given in Eq.~(\ref{eom.met}) can be used to find the corrections about known backgrounds of GR. Such corrections about the Schwarzschild and FRW backgrounds were described in~\cite{Maggiore:2014sia}, based on the analysis carried out for Eq.~(\ref{eom.mag1}) in \cite{Kehagias:2014sda}. We will now briefly review this derivation for the Schwarzschild background. By considering the following spherically symmetric $4 d$ metric 
\begin{equation}
ds^2 = -e^{2 \alpha(r)} dt^2 + e^{2 \beta(r)} dr^2 + r^2 (d\theta^2 + \text{sin}^2 \theta d\phi^2) \, ,
\label{met.an1}
\end{equation}
its substitution in Eq.~(\ref{mag.aux}) provides the following two equations
\begin{align}
r^2 U'' + \left[2r+ (\alpha' - \beta')r^2\right] U' = -2 e^{2\beta} + 2 &\left[1 + 2r(\alpha' - \beta') + r^2 (\alpha'' + \alpha'^2 - \alpha' \beta')\right] \, , \label{eq1} \\
S'' + (\alpha' - \beta' + \frac{2}{r})S' &= e^{2\beta} U \, .\label{eq2}
\end{align}
Likewise, from Eq.~(\ref{eom.met}) and Eq.~(\ref{met.an1}) the equations for $e^{2(\beta - \alpha)}R_{00} + R_{11}$ and $R_{22}$ take the following form
\begin{align}
(1 - 2\mu S) (\alpha' + \beta') &= -\mu r \left[ S'' - (\alpha' +\beta' - U') S' \right] \, , \label{eq3}\\
(1 - 2\mu S) \left[1 + e^{-2\beta}(r(\beta' - \alpha') - 1) \right] & = \mu \left[r^2 \left(U + \frac{U^2}{2}\right) - 2 r e^{-2\beta} S' \right] \, . \label{eq4}
\end{align}
Eqs.~(\ref{eq1}-\ref{eq4}) represent four independent equations in four unknowns. 
The primes in these equations denote differentiation with respect to $r$, i.e. $\alpha' = \frac{\partial \alpha}{\partial r}$. Due to the non-vanishing $U$ and $S$ fields, the right hand side of Eq.~(\ref{eq3}) does not vanish as it does in GR and leads to $\alpha$ being in general different from $\beta$. The complete solutions were not derived for these coupled differential equations. However they can be solved in terms of corrections about the Schwarzschild background in the region far away from the black hole. 

In the Newtonian limit where $r_S \ll r$, with $r_S$ denoting the Schwarzschild radius, we can consider perturbations about flat space and $m$ arbitrary. The solution for $U(r)$ resulting from Eq.~(\ref{eq1}) is now given by the Green function for the inhomogeneous Helmholtz equation. With this solution for $U$, the following expressions for $A(r) (= e^{2\alpha})$ and $B(r) (= e^{2 \beta})$ were derived
\begin{align}
A(r) &=  1 - \frac{r_S}{r}\left[1 + \frac{1}{3}\left(1 - \text{cos}\,mr\right) \right] \, , \notag\\
B(r) &= 1 + \frac{r_S}{r}\left[1 - \frac{1}{3}\left(1 - \text{cos}\,mr - mr\,\text{sin}\,mr\right) \right] \, ,
\label{sol.A2}
\end{align} 

The corrections can also be derived in the small $m$ limit, i.e. $r \ll m^{-1}$. In this limit, the metric approximates to those of the Schwarzschild background, i.e. $\alpha  \approx \text{ln}(1 - \frac{r_S}{r}) \approx  -\beta$. Eq.~(\ref{eq1}) then simplifies to $\Box U = 0$ about the Schwarzschild background, which has the following general solution
%
%
\begin{equation}
U(r) = u_0 - u_1 \text{ln}(1 - \frac{r_S}{r}) \, .
\label{eom.U1}
\end{equation}
This expression can be substituted in Eqs.~(\ref{eq2}, \ref{eq3}, \ref{eq4}), to solve for $A(r)\,,B(r)$ and $S(r)$. For real constants $u_0$ and $u_1$, the solution of Eq.~(\ref{eom.U1}) is real for all values of $r$ beyond the Schwarzschild radius. The constant $u_0$ however provides a cosmological constant to the field equations. This can be noted from Eq.~(\ref{eom.met}), where the $\mu \frac{U^2}{2} g_{\mu \nu}$ term contained in $K_{\mu \nu}$ would provide such a contribution. Since we are considering corrections about the Schwarzschild background, this constant should be set to vanish. The corrections resulting from Eq.~(\ref{eom.U1}) were found to agree with the corresponding expressions in Eq.~(\ref{sol.A2}) provided $u_1=1$.
With these values for the constants, we have
\begin{equation}
U(r) = -\text{ln}(1 - \frac{r_S}{r}) \, .
\label{eom.U2}
\end{equation}
Note that while this solution is not well defined for $r \le r_S$ (near the horizon), it does nevertheless allow for corrections outside the horizon of the black hole. Eq.~(\ref{eom.U2}) provides the following leading order correction to the metric when $r_S \ll r$
\begin{equation}
A(r) = e^{2 \alpha} \approx 1 - \frac{r_S}{r}\left(1+ \mu r^2\right) \, .
\label{metric.corr}
\end{equation}
The solutions were further considered via numerical integration to account for corrections beyond first-order. This verified that the corrections to GR remain $1+\mathcal{O}(m^2r^2)$ and linear up to $r \sim r_S$ (since $m \sim H_0$) at higher orders. Thus one recovers the vacuum solution of Einstein's equations in the limit of $m \to 0$, demonstrating that the theory contains no vDVZ discontinuity. This is in contrast with the result in the case of the Einstein-Hilbert action with a Fierz-Pauli term, where a vDVZ disconuity does result. The linear expansion breaks down below a Vainshtein radius  $r_V = (\frac{GM}{m^4})^{1/5}$, which is parametrically larger than the Schwarzschild radius.

\subsection{Corrections about an electrovacuum background}
In this subsection, we will briefly describe how the minimally coupled Maxwell action does not lead to real corrections outside the event horizon of a RN black hole. Let us consider the inclusion of the Maxwell action 
%
\begin{equation}
S_{M} = - \frac{1}{16 \pi} \int d^4 x \sqrt{-g}  F^{\beta \gamma} F_{\beta \gamma} \,,
\label{max.act}
\end{equation}
in Eq.~(\ref{mag.act}). The gravitational field equations are as given in Eq.~(\ref{eom.met}) and from Eq.~(\ref{eom.set}) we have the following stress-energy tensor
\begin{equation}
T_{\mu \nu} = \frac{1}{4 \pi} \left( F_{\mu \alpha} F_{\nu}^{\phantom{\nu} \alpha} - \frac{1}{4} F^{\beta \gamma} F_{\beta \gamma} g_{\mu \nu} \right) \,.
\label{max.set}
\end{equation}
Varying with respect to the electromagnetic field provides Maxwell's equations, which are not modified in the present case
\begin{equation}
\nabla_{\mu}F^{\mu \nu} = 0 \label{max.max}
\end{equation}
As in the case of the Schwarzschild background, we can solve Eq.~(\ref{eom.met}) by assuming the spherically symmetric metric given in Eq.~(\ref{met.an1}).
Without assuming any particular ansatz for the electromagnetic field, we can now show that real perturbative corrections do not exist outside the event horizon of a charged black hole solution of GR, such as the RN black hole.  Denoting the mass and charge of the RN black hole by $M$ and $Q$ respectively, the small $m$ limit in this case implies $\alpha \approx \text{ln}(1 - \frac{2M}{r} + \frac{Q^2}{r^2}) \approx  -\beta$.  We denote the horizon radius in this case as $r_H = M + \sqrt{M^2 - Q^2}$. The constraint equation in Eq.~(\ref{eq1}) now becomes $\Box U = 0$ about the RN background, whose general solution is given by
\begin{equation}
U = c_1 + \frac{c_2}{\sqrt{M^2 - Q^2}} \text{ArcTanh}\left(\frac{r-M}{\sqrt{M^2 - Q^2}}\right)  \, .
\label{sol.tanh}
\end{equation}
$\text{ArcTanh}(x)$ is real only when its argument lies between $x \in [-1,1]$. In the context of its argument in Eq.~(\ref{sol.tanh}), this region lies between $M-\sqrt{M^2 - Q^2}$ and $M+\sqrt{M^2 - Q^2}$. Thus for real coefficients $c_1$ and $c_2$, real solutions exist only in the region between the inner Cauchy horizon and the event horizon of the RN black hole. The region beyond the event horizon can be only considered through complex coefficients.
As an interesting point, we also note that in the limit of $Q \to 0$ one does not recover the Schwarzschild limit result Eq.~(\ref{eom.U2}). 
%
%
%
%
Unlike Eq.~(\ref{eom.U1}), the solution given in Eq.~(\ref{sol.tanh}) does not admit a real, small $m$ correction outside the event horizon of the RN black hole.  Thus real solutions of Eq.~(\ref{eom.met}) with an electromagnetic stress-energy tensor cannot be perturbative about the external RN background. In a means to resolve this, we will consider the non-minimal coupling of the electromagnetic field to the theory via the Kaluza Ansatz in the next section. 
 
\section{The Kaluza Reduced action} \label{sec.kk}
%
In this section we will describe the Kaluza ansatz, which can be used to geometrically determine the electromagnetic coupling to a given pure gravitational theory. We assume a five-dimensional spacetime with the following metric
\begin{equation}
\hat{g}_{AB} = 
\begin{pmatrix}
g_{\mu\nu} + {\alpha}^2 A_{\mu}A_{\nu}  &  {\alpha} A_{\mu} \\ 
{\alpha} A_{\nu} &  1 
\end{pmatrix}\,,
\label{kk.met}
\end{equation}
where ${\alpha}$ is a parameter which will be fixed later. Here and elsewhere in this section, five-dimensional objects will be represented with hats, uppercase Latin indices are five-dimensional, $A,B,\dots = 0,\cdots,4,$ while Greek indices are four-dimensional, $\mu,\nu,\dots = 0,\cdots,3$\,. The inverse of Eq.~(\ref{kk.met}) is given by 
\begin{equation}
\hat{g}^{AB} = 
\begin{pmatrix}
g^{\mu\nu}  & - {\alpha} A^{\mu} \\
- {\alpha} A^{\nu} &  {\alpha}^2 A^{\gamma} A_{\gamma} + \frac{1}{{\Phi}^2} 
\end{pmatrix}\,.
\label{kk.minv}
\end{equation}
From Eq.~(\ref{kk.met}), it follows that $\sqrt{-\hat{g}} = \sqrt{-g}$.  We will further assume the cylindricity condition $\frac{\partial \hat{g}_{AB}}{\partial x^5} = 0$. An immediate consequence of this condition is that $\hat{\Box} = \hat{g}^{AB}\hat{\nabla}_{A}\hat{\nabla}_{B} = g^{\mu \nu} \nabla_{\mu}\nabla_{\nu} = \Box$. Using the metric given in Eq.~(\ref{kk.met}), we find the following components of the Ricci tensor
\begin{align}
\hat{R}_{\mu\nu} &= R_{\mu\nu} + \frac{1}{4}{\alpha}^4  F^{\beta\gamma}F_{\beta\gamma} A_{\mu}A_{\nu} -\frac{1}{2}{\alpha}^2 (A_{\mu}{\nabla}_{\beta} F^{\beta}_{\phantom{\beta}\nu} + A_{\nu}{\nabla}_{\beta} F^{\beta}_{\phantom{\beta}\mu} + F_{\beta\mu}F^{\beta}_{\phantom{\beta} \nu}) \, , \notag \\
\hat{R}_{\mu 5} &= \frac{1}{4}{\alpha}^3 F^{\beta \gamma}F_{\beta  \gamma} A_{\mu} - \frac{1}{2}{\alpha}({\nabla}_{\beta}F^{\beta}_{\phantom{\beta}\mu})\,, \qquad
\hat{R}_{55} = \frac{1}{4}{\alpha}^2 F^{\beta \gamma} F_{\beta  \gamma}\,.
\label{kk.riccs}
\end{align}
The Ricci tensor components can now be used to derive the five-dimensional Ricci scalar
\begin{equation}
\hat{R} = R - \frac{{\alpha}^2}{4} F^{\beta \gamma} F_{\beta  \gamma}\,.
\label{kk.scal}
\end{equation}
Writing the radius of compactification of the fifth dimension as $\tilde R\,,$ and the five-dimensional Newton's constant as $\hat G_5\,,$ we find that setting 
\begin{equation}
 \frac{2 \pi \tilde{R}}{{\hat{G}}_{5}} = \frac{1}{G} = 1 \, , \qquad {\alpha}^2 = 4G = 4 \, ,
\label{kk.const}
\end{equation}
leads to the following reduction of the five-dimensional Einstein-Hilbert action 
\begin{align}
\frac{1}{16 \pi {\hat{G}}_{5}} \int d^5x \sqrt{- \hat{g}} \hat{R} = \frac{1}{16 \pi} \int d^4x \sqrt{-g} R -\frac{1}{16 \pi}\int d^4x \sqrt{-g} F^{\beta \gamma} F_{\beta \gamma} \,.
\label{kk.EH}
\end{align}
~~~Thus the choice considered in Eq.~(\ref{kk.const}) leads to the usual Einstein-Maxwell action in four dimensions. Let us now consider Eq.~(\ref{mag.act1}) in five dimensions. The action in this case is given by
\begin{equation}
\hat{S}_{MM} =  \frac{1}{16 \pi \hat{G}_5}\int d^5x \sqrt{-\hat{g}} \left[\hat{R} - \hat{\mu} \hat{R} \frac{1}{\hat{\Box}^2}\hat{R} \right] \, ,
\label{kk.mact1}
\end{equation}
The five-dimensional mass scale $\hat{\mu}$ must be such that $\hat{\mu} = \frac{\hat{m}^2}{6} = \frac{m^2}{6} = \mu$, as the mass scale must match the original non-local theory in the limit of a vanishing electromagnetic field. Likewise, since $\hat{\Box}^2 = \Box^2$, their inverses should also agree. For the remaining terms, we substitute Eq.~(\ref{kk.const}) and Eq.~(\ref{kk.scal}) to find the following reduced action   
\begin{equation}  
S_{KMM} = \frac{1}{16 \pi} \int d^4x \sqrt{-g} \left[R -  F^{\beta \gamma} F_{\beta \gamma} - \mu \left((R -  F^{\beta \gamma} F_{\beta \gamma}) \frac{1}{\Box^2} (R -  F^{\beta \gamma} F_{\beta \gamma})\right)\right]
\label{kk.mact2}
\end{equation}
Eq.~(\ref{kk.mact2}) now comprises non-local terms which involve both the Ricci scalar and the electromagnetic field strength tensor. 
In the next section, we will consider the local formulation of this action through the introduction of auxilliary fields.
The constraint equations satisfied by these auxilliary fields will play an important role in the nature of the resulting solutions.

\section{Equations of motion}\label{sec.eom}
Let us define two variables $\tilde{U}$ and $\tilde{S}$ which satisfy
\begin{equation}
\Box \tilde{U}= - \left(R - F^{\beta \gamma} F_{\beta \gamma}\right) \, , \quad  \Box \tilde{S}= - \tilde{U} \, .
\label{eom.corr2}
\end{equation}
The presence of $F^{\beta \gamma} F_{\beta \gamma}$ as a source for the $\tilde{U}$ field distinguishes the fields given in Eq.~(\ref{eom.corr2}) with those of Eq.~(\ref{mag.aux}). The fields $\tilde{U}$ and $\tilde{S}$ can now be substituted in Eq.~(\ref{kk.mact2}) with the help of two Lagrange multipliers $\xi_1$ and $\xi_2$ to give
\begin{equation}
S_{KMM} = \frac{1}{16 \pi} \int d^4x \sqrt{-g} \left[(R-  F^{\beta \gamma} F_{\beta \gamma})(1 - \mu \tilde{S}) - \xi_1\left(\Box \tilde{U} + R -  F^{\beta \gamma} F_{\beta \gamma}\right) - \xi_2\left(\Box\tilde{S} + \tilde{U}\right) \right] \, .
\label{eom.meom}
\end{equation}
Varying this action with respect to $\tilde{U}$ and $\tilde{S}$ we find
\begin{align}
-\xi_2 - \Box \xi_1 &= 0 \, , \notag\\
-\mu (R -  F^{\beta \gamma} F_{\beta \gamma}) - \Box \xi_2 &= 0 \, ,
\end{align}
respectively. These two equations identify $\xi_2= \mu \tilde{U}$ and $\xi_1 = \mu \tilde{S}$. From Eq.~(\ref{eom.meom}) we find the following equation of motion for the metric
\begin{equation}
\left(G_{\mu \nu} - 8 \pi T_{\mu \nu}\right) \left(1 - 2\mu \tilde{S}\right) = \mu \tilde{K}_{\mu \nu} \, ,
\label{eom.meom2}
\end{equation}
where $T_{\mu \nu}$ is as defined in Eq.~(\ref{max.set}) and
\begin{equation}
\tilde{K}_{\mu \nu} = g_{\mu \nu} \left(2~ \Box \tilde{S} + \nabla_{\alpha}\tilde{U}\nabla^{\alpha}\tilde{S} - \frac{1}{2}\tilde{U}^2 \right)  - 2 \nabla_{\mu} \nabla_{\nu}\tilde{S} - \left(\nabla_{\mu}\tilde{U} \nabla_{\nu}\tilde{S} + \nabla_{\nu}\tilde{U} \nabla_{\mu}\tilde{S}\right) \, .
\label{eom.mcorr}
\end{equation}
The equations of motion for the electromagnetic field $A_{\mu}$ resulting from Eq.~(\ref{eom.meom}) is given by
\begin{equation}
\nabla_{\mu}\left(\left(1 - 2\mu \tilde{S}\right)F^{\mu \nu}\right)=0 \, .
\label{eom.fcorr}
\end{equation}
Thus both Einstein's and Maxwell's equations involve non-local corrections. This is unlike the situation where the electromagnetic field is minimally coupled.
%
 In comparing Eq.~(\ref{eom.meom2}) with Eq.~(\ref{eom.met}) we note that the stress-energy tensor in the Kaluza reduced case also involves $\mu$ corrections. Finally, while Eq.~(\ref{eom.mcorr}) and Eq.~(\ref{eom.corr}) show that the general correction terms $\tilde{K}_{\mu \nu}$ and $K_{\mu \nu}$ are structurally similar, they provide different contributions due to the difference in the definitions of the auxilliary fields. In the following subsection, we will derive the solution of Eq.~(\ref{eom.meom2}) through an iterative approach built on known solutions of GR.

\section{Iterative approach for the solutions}\label{sec.sol}
To construct the solutions of the equations of motion given in the previous section, let us rewrite Eq.~(\ref{eom.meom2}) and Eq.~(\ref{eom.fcorr}) in the following way
\begin{align}
G_{\mu \nu} - 8 \pi T_{\mu \nu} &= \mu \left(\tilde{K}_{\mu \nu} + 2 \tilde{S} (G_{\mu \nu} - 8 \pi T_{\mu \nu})\right) \, \notag\\
\nabla_{\mu} F^{\mu \nu} &= 2 \mu \left((\nabla_{\mu}\tilde{S})F^{\mu \nu} + \tilde{S}\nabla_{\mu}F^{\mu \nu}\right) \, .
\label{sol.eq}
\end{align}
The form of these equations suggest that we can consider the fields $g_{\mu\nu}$ and $A_\mu$ in terms of their zeroth-order and first-order (in $\mu$) contributions 
\begin{equation}
g_{\mu\nu} = g^{(0)}_{\mu\nu} +  g^{(1)}_{\mu\nu}\,, \qquad 
A_\mu = A^{(0)}_\mu + A^{(1)}_\mu\,.
\label{sol.lo}
\end{equation}
Here $\{g^{(1)}_{\mu\nu}\, , A^{(1)}_\mu\}$ is linear in $\mu$ and $\{g^{(0)}_{\mu\nu}\,,A^{(0)}_\mu\}$ satisfy the the Einstein-Maxwell equations 
\begin{equation}
G^{(0)}_{\mu \nu} - 8 \pi T^{(0)}_{\mu \nu} =0 \,, \qquad \nabla^{(0)}_{\mu} F^{(0) \mu \nu} = 0 \, .
\label{sol.rn}
\end{equation}
Thus $g^{(0)}_{\mu\nu}$ and $A^{(0)}_\mu$ represent any known electrovacuum solution of GR. Here we will assume the standard RN solution
\begin{equation}
g^{(0)}_{\mu\nu} = 
\begin{pmatrix}
- f(r)  & 0 & 0 & 0 \\ 
0 & {f(r)}^{-1} & 0 & 0\\
0 & 0 & r^2 &  0\\
0 & 0 & 0 & r^2 {\text{sin}}^{2} {\theta}\\
\end{pmatrix}\, , \qquad
F^{(0)}_{\mu\nu} = 
\begin{pmatrix}
0 & \frac{Q}{r^2} & 0 & 0 \\ 
- \frac{Q}{r^2} & 0 & 0 & 0\\
0 & 0 & 0 &  0\\
0 & 0 & 0 & 0\\
\end{pmatrix}\, ,
\label{met.rn}
\end{equation}
where $f(r) = 1 - \frac{2M}{r} + \frac{Q^2}{r^2}\,$, with $M$ and $Q$ denoting the mass and charge of the black hole respectively, while $F^{(0)}_{\mu \nu} = 2 \partial_{[\mu}A^{(0)}_{\nu]}$ is the lowest order electromagnetic field strength tensor. Using these solutions, we can find the first-order corrections in $\mu$ from Eq.~(\ref{sol.eq})   
\begin{align}
G^{(1)}_{\mu \nu} - 8 \pi T^{(1)}_{\mu \nu} &= \mu \left(\tilde{K}^{(0)}_{\mu \nu} + 2 \tilde{S}^{(0)} (G^{(0)}_{\mu \nu} - 8 \pi T^{(0)}_{\mu \nu})\right) = \mu \tilde{K}^{(0)}_{\mu \nu} \,, \notag\\
\nabla^{(1)}_{\mu} F^{(0) \mu \nu} + \nabla^{(0)}_{\mu} F^{(1) \mu \nu} &= 2 \mu \left((\nabla^{(0)}_{\mu}\tilde{S}^{(0)})F^{(0) \mu \nu} + \tilde{S}^{(0)}\nabla^{(0)}_{\mu}F^{(0) \mu \nu}\right) = 2 \mu (\nabla^{(0)}_{\mu}\tilde{S}^{(0)})F^{(0) \mu \nu} \, .
\label{sol.eq2}
\end{align}
In Eq.~(\ref{sol.eq2}), $\nabla^{(0)}_{\mu}$ and $\nabla^{(1)}_{\mu}$ imply that the connection in the covariant derivative involve terms up to the respective order. $G^{(1)}_{\mu\nu}$ and $T^{(1)}_{\mu\nu}$ are the first-order contributions of $G_{\mu\nu}$ and $T_{\mu\nu}\,,$ while $\tilde{K}^{(0)}_{\mu \nu}$ is described in terms of the zeroth-order fields with the following expression 
\begin{align}
\tilde{K}^{(0)}_{\mu \nu} &= g^{(0)}_{\mu \nu} \left(2 \Box^{(0)} \tilde{S}^{(0)} + \nabla_{\alpha}^{(0)}\tilde{U}^{(0)}\nabla^{(0) \alpha}\tilde{S}^{(0)} - \frac{1}{2}\tilde{U}^{(0) 2} \right)  - 2 \nabla_{\mu}^{(0)} \nabla^{(0)}_{\nu}\tilde{S}^{(0)} \notag\\
&\qquad \qquad \qquad  - \left(\nabla^{(0)}_{\mu}\tilde{U}^{(0)} \nabla^{(0)}_{\nu}\tilde{S}^{(0)} + \nabla^{(0)}_{\nu}\tilde{U}^{(0)} \nabla^{(0)}_{\mu}\tilde{S}^{(0)}\right) \,.
\end{align}
To solve Eq.~(\ref{sol.eq2}), we will assume the following ansatz for the spherically symmetric metric and electromagnetic field strength
\begin{align}
g_{\mu\nu} &= 
\begin{pmatrix}
- (f(r) + \mu A(r))  & 0 & 0 & 0 \\ 
0 & {(f(r) + \mu(A(r) - B(r)))}^{-1} & 0 & 0\\
0 & 0 & r^2 &  0\\
0 & 0 & 0 & r^2 {\text{sin}}^{2} {\theta}\\
\end{pmatrix}\, , \notag\\
F_{\mu\nu} &= 
\begin{pmatrix}
0 & \frac{Q}{r^2}+ \mu D(r) & 0 & 0 \\ 
- \frac{Q}{r^2}-\mu D(r) & 0 & 0 & 0\\
0 & 0 & 0 &  0\\
0 & 0 & 0 & 0\\
\end{pmatrix}\, ,
\label{met.rn2}
\end{align}
where $A(r)\,, B(r)$ and $D(r)$ represent the first-order correction terms. Let us first consider the constraint equations given in Eq.~(\ref{eom.corr2}), which will be needed to solve Eq.~(\ref{sol.eq2}). The constraint equations are satisfied at all orders in $\mu$. The first-order corrections to the metric and electromagnetic field strength tensor require the zeroth-order solutions of the constraint equations. From Eq.~(\ref{eom.corr2}), we have the following lowest order equations
\begin{equation}
\Box^{(0)} \tilde{U}^{(0)} = F^{(0)}_{\alpha \beta}F^{(0)\alpha \beta} = -\frac{2 Q^2}{r^4} \, , \qquad \Box^{(0)} \tilde{S}^{(0)} = - \tilde{U}^{(0)} \, .
\label{sol.us}
\end{equation}
To simplify the notation, we will henceforth label $\tilde{U}^{(0)}(r)$ and $\tilde{S}^{(0)}(r)$ as $\tilde{u}(r)$ and $\tilde{s}(r)$ respectively. Using this notation, from Eq.~(\ref{met.rn}) and Eq.~(\ref{sol.us}) we have the following two equations about the RN background
\begin{align}
\frac{2 Q^2}{r^4} & = - f(r)'\tilde{u}(r)'+ f(r) \left(\frac{2\tilde{u}(r)'}{r} + \tilde{u}(r)'' \right) \label{sol.ut} \\
\tilde{u}(r) & = f(r)'\tilde{s}(r)'+ f(r) \left(\frac{2\tilde{s}(r)'}{r} + \tilde{s}(r)'' \right) \, ,  \label{sol.st} 
\end{align}
where primes denote differentiation with respect to coordinate `$r$'.
The first-order correction of the $(1-2\mu S) (g^{00}R_{00} -g^{11}R_{11})$ term can be found from the first equation of Eq.~(\ref{sol.eq2})
\begin{align}
f(r) B(r)' - B(r) f(r)' &=- 2 r f(r)^2(\tilde{u}(r)'\tilde{s}(r)'+ \tilde{s}(r)'') \, .\label{sol.B}
\end{align}
Likewise, the second equation of Eq.~(\ref{sol.eq2}) provides the first-order correction to Maxwell's equation
\begin{align}
Q \left(B(r)f(r)' - B(r)'f(r)\right) &= -2 f(r)^2\left(- Q \tilde{s}(r)'+ 2 r D (r) + r^2 D(r)'\right) \, . \label{sol.D}
\end{align}
Finally, the first-order correction to $R_{22}$ can also be determined from the first equation of Eq.~(\ref{sol.eq2}) to be
\begin{align}
4r f(r) \tilde{s}(r)' &= 2\left(A(r) + 2 Q D(r) + r^2\left(\tilde{u}(r) + \frac{\tilde{u}(r)^2}{2}\right) + r\left(A(r)' - \frac{B(r)'}{2}\right)\right) \notag\\
&\qquad \qquad - B(r)\left(2+\frac{2Q^2}{f(r) r^2} + \frac{r f(r)'}{f(r)}\right) \, .\label{sol.A}
\end{align}
Eqs.~(\ref{sol.ut}-\ref{sol.A}) provide a sequence of equations which can be used to solve the coupled differential equations. Beginning with 
Eq.~(\ref{sol.ut}), we can find a solution for $\tilde{u}(r)$. Using this solution in Eq.~(\ref{sol.st}), we can solve for $\tilde{s}(r)$.  These two solutions can be used in Eq.~(\ref{sol.B}), where we now solve for the field $B(r)$. Proceeding in this way, we can solve for the fields $\tilde{u}(r)\, , \tilde{s}(r)\, ,B(r)\,, D(r)\,$ and $A(r)$ sequentially and determine the first-order correction to the field equations.
This sequence also demonstrates the pivotal role of $\tilde{u}$ in providing the corrections.
\subsection{First order corrections}
The general solution of Eq.~(\ref{sol.ut}) on the RN background is given by
\begin{equation}
\tilde{u} = c_1 + \frac{c_2}{\sqrt{M^2 - Q^2}} \text{ArcTanh}\left( \frac{r-M}{\sqrt{M^2 - Q^2}}\right) - \text{ln}\left(1 - \frac{2M}{r} + \frac{Q^2}{r^2} \right)  \, .
\label{sol.log}
\end{equation}
Following the discussion below Eq.~(\ref{sol.tanh}), there are no real constants $c_1$ and $c_2$ which admit real corrections outside the event horizon of a RN black hole. By setting $c_1 = 0 = c_2$, the homogeneous solution can be eliminated and we are left with 
\begin{equation}
\tilde{u} = - \text{ln}\left(1 - \frac{2M}{r} + \frac{Q^2}{r^2} \right)  \, .
\label{sol.log2}
\end{equation}
This solution, apart from being real outside the event horizon of the RN black hole, also agrees with Eq.~(\ref{eom.U2}) in the limit of vanishing charge. Neither Eq.~(\ref{eom.U2}) nor Eq.~(\ref{sol.log2}) are real for $r \le r_H$, due to the fact that both solutions of the constraint equations were derived in static coordinates. We could always derive a solution for $\tilde{u}$ by adopting coordinates which are well defined across the horizon. However, as we intend to derive the perturbative corrections outside the RN black hole and compare these with the known corrections about the Schwarzschild background, we will continue to adopt the usual spherically symmetric coordinates in this section. 

The solution of $\tilde{u}$ is real outside the event horizon of a RN black hole and thus allows us to consider perturbative corrections in $\mu$ in this region.
Using Eq.~(\ref{sol.log2}) in Eq.~(\ref{sol.st}), the general solution of $\tilde{s}$ comprises of logarithmic terms, products of logarithmic terms and polylog functions, which are similar to those found about the Schwarzschild background \cite{Kehagias:2014sda}. Further, unlike the Schwarzschild case, there exist the homogeneous contribution involving the $\text{ArcTanh}\left( \frac{r-M}{\sqrt{M^2 - Q^2}}\right)$ functions, which can again be addressed through an appropriate choice of the constants. The solutions for $A(r)\, , B(r)$ and $D(r)$ are significantly more involved in the logarithmic terms which can obscure their leading order behaviour when $r_H \ll r$. We will thus consider only the leading order in $r$ contribution for all the correction terms, for which we find the following 
\begin{align}
A(r) &= -2 M r - 2(M^2 - Q^2)  + d_1 + \mathcal{O}(r^{-1}) \, \notag\\
B(r) &= d_1 + \mathcal{O}(r^{-1}) \notag\\
D(r) &= -\frac{2 M Q}{r} + \mathcal{O}(r^{-2}) \,
\label{sol.kkcor}
\end{align}
The lowest order correction to the RN background can be found by taking $d_1 = 0$. Substituting Eq.~(\ref{sol.kkcor}) in Eq.~(\ref{met.rn2}) we find the following non-vanishing corrections of the metric and electric field, to first-order in $\mu$ and leading orders in $r$
\begin{align}
g_{00} &\approx \left(1 - \left( \frac{2M}{r} \right)(1 + \mu (r^2 + M r)) +   \frac{Q^2}{r^2}(1 + 2 \mu r^2) \right)  \notag\\
F_{01} & \approx \frac{Q}{r^2}\left(1 - 2 \mu M r \right) 
\label{sol.mete}
\end{align}
\begin{figure}
\begin{center}
\includegraphics[width=0.42\columnwidth]{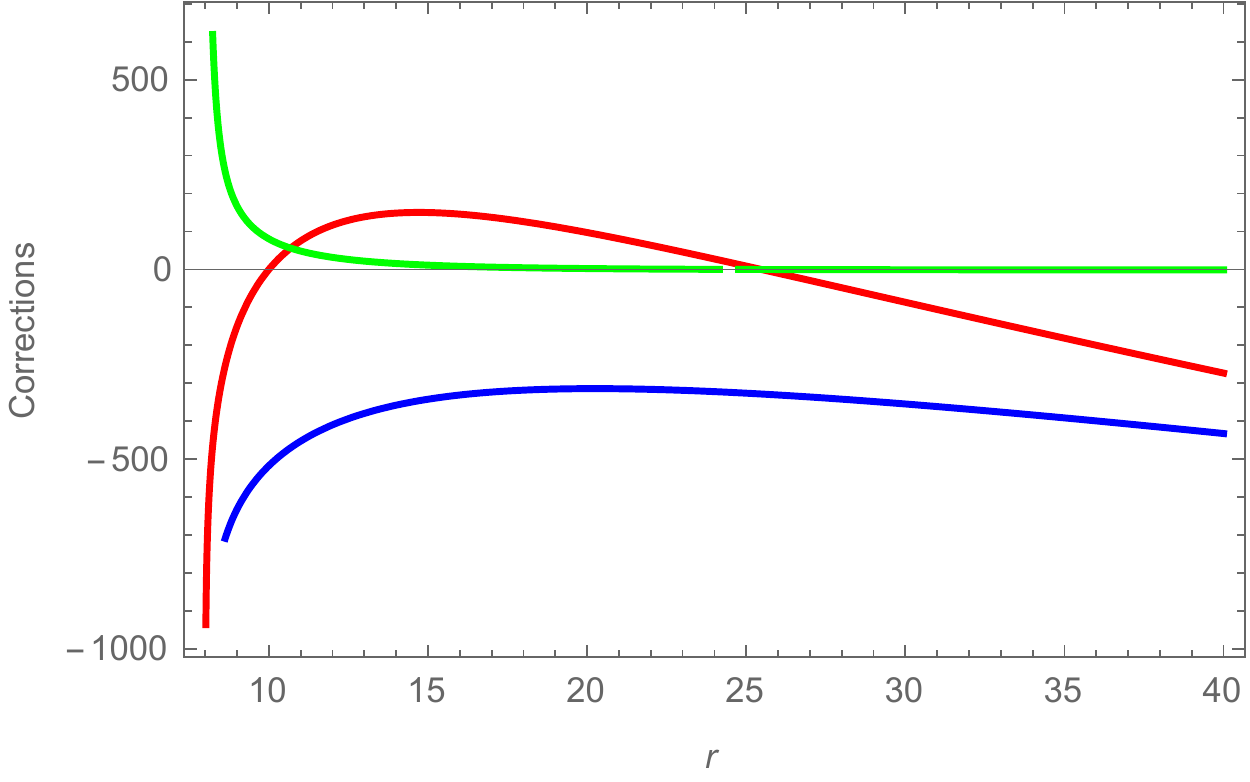}
\includegraphics[width=0.44\columnwidth]{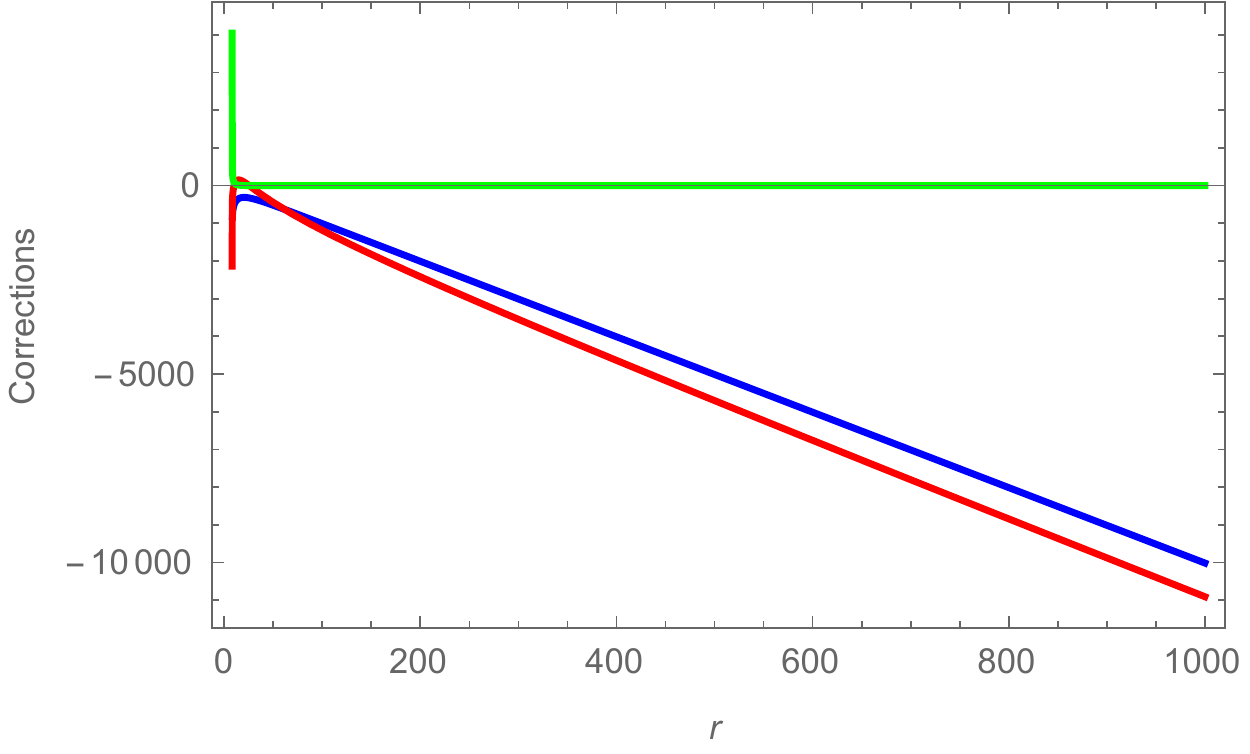}
\caption{In all plots $M=5$ and $Q=4$ are assumed; The event horizon is located at $r_H=8$. \newline Left: The complete first-order correction funtions $A(r)$ (red), $B(r)$ (blue) and $D(r)$ (green) near the horizon of the black hole (From $r = 8$ to $r = 40$). \newline Right: The same functions now considered up to r=1000.  $A(r) \, ,B(r) \,$ and $D(r)$ approach finite values at the horizon and are well behaved outside the horizon.}
\end{center}
\end{figure}
As can be seen from comparing Eq.~(\ref{metric.corr}) with Eq.~(\ref{sol.mete}), the corrected metric is an appropriate extension of the result for the Schwarzschild background. We also note that the electric field contains a charge correction which involves the mass $M$ of the black hole. Such $r^{-1}$ corrections for the electric field are absent when the electromagnetic field is minimally coupled. The corrections beyond leading order in Eq.~(\ref{sol.kkcor}) could be relevant in the region beyond the horizon. We have thus also taken into account the complete solution for $A(r)\, , B(r)$ and $D(r)$ numerically. The plots for these solutions are provided in Figure 1. These should not be confused with $\mu A(r) \, , \mu B(r)$ and $\mu D(r)$, which provide the actual corrections to the metric and electromagnetic field strength and are neglibly small except for $mr \ge 1$.  
The corrected metric and field strength have been compared with the uncorrected RN solutions in Figures 2 and 3. These figures indicates the excellent agreement of the complete first-order corrected solution with the original RN solution from $r \approx r_H$ up to $mr < 1$.

\begin{figure}
\begin{center}
\includegraphics[width=0.43\columnwidth]{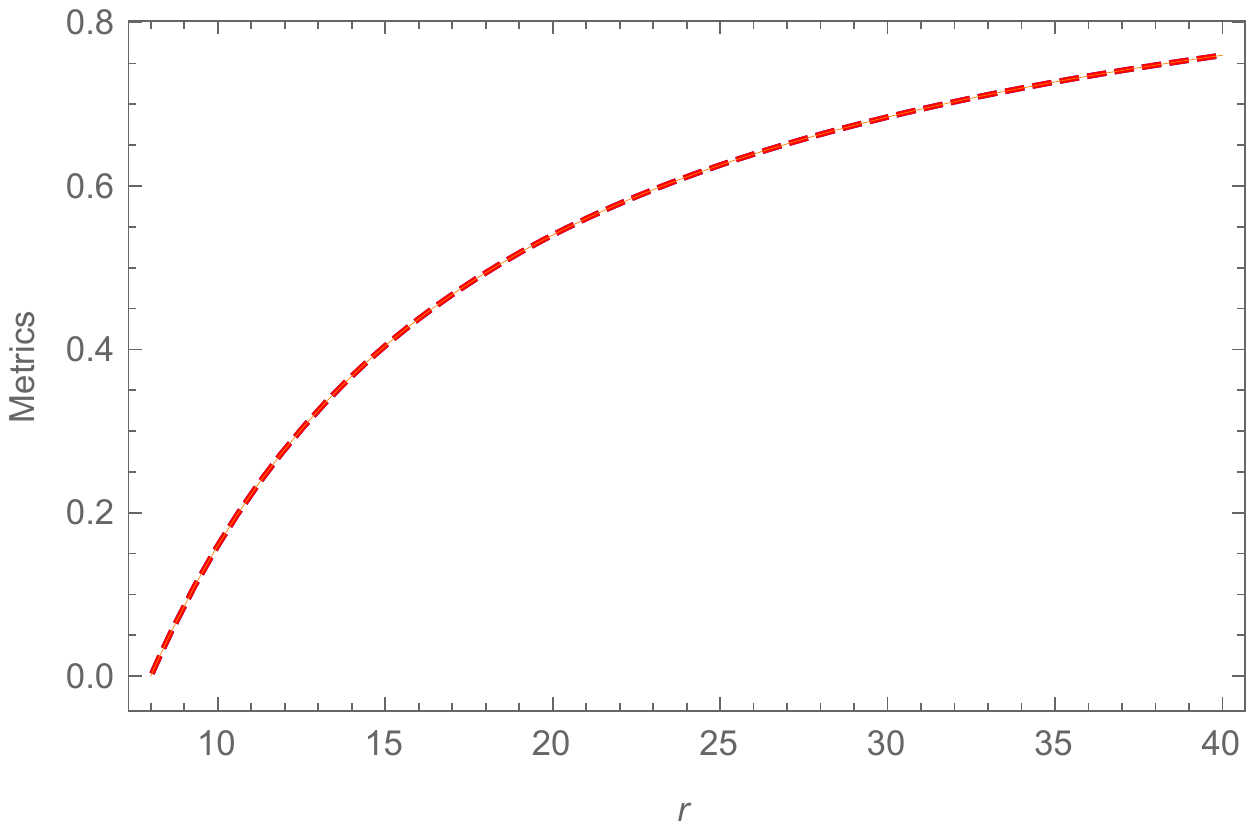}
\includegraphics[width=0.45\columnwidth]{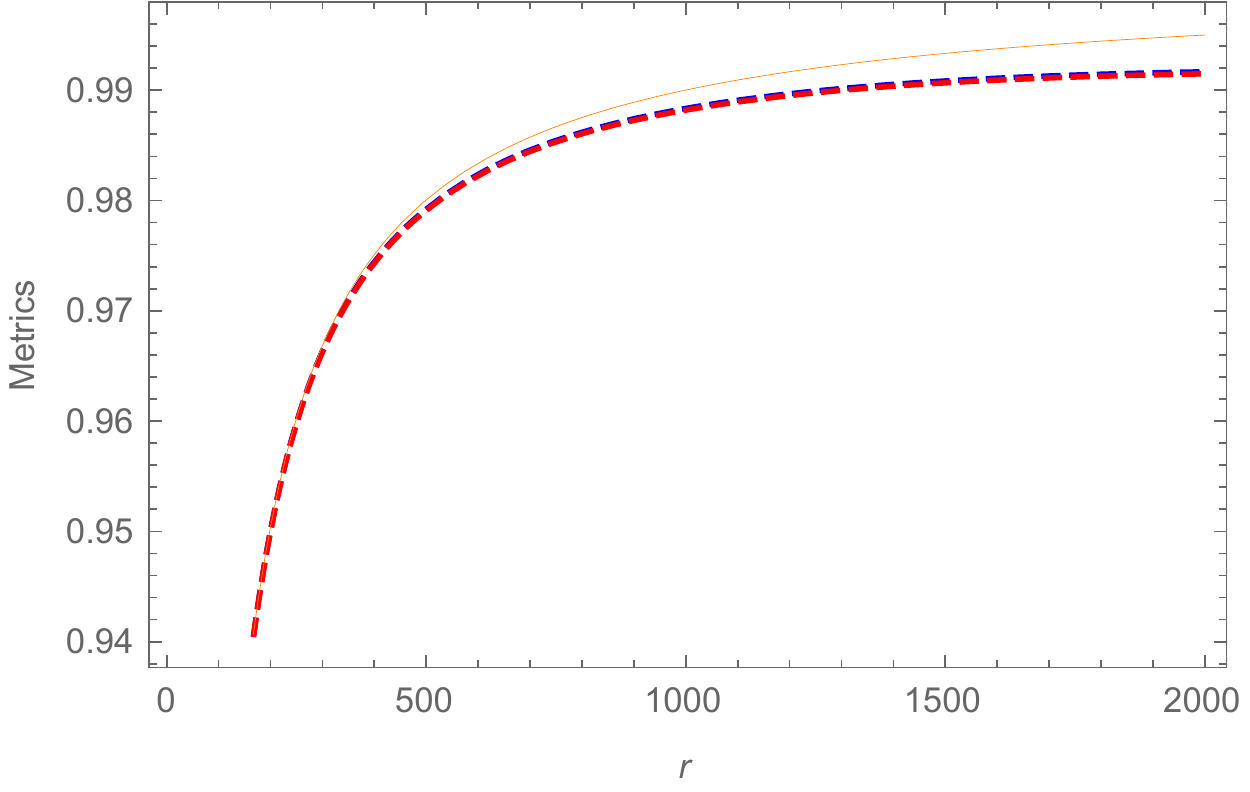}
\caption{Comparative plots of the complete first-order corrected $g_{00}$ (red dashed) and $g^{11}$ (blue dashed) metric components with that of the uncorrected RN $g^{(0)}_{00}$ (yellow) assuming $M = 5\, , Q = 4$ and $\mu = \frac{m^2}{6} = 10^{-6}$.  \newline Left: From  $r=8$ to $r=40$. ~~~ Right: From  $r=8$ to $r=2000$}
\end{center}
\end{figure}
\begin{figure}
\begin{center}
\includegraphics[width=0.43\columnwidth]{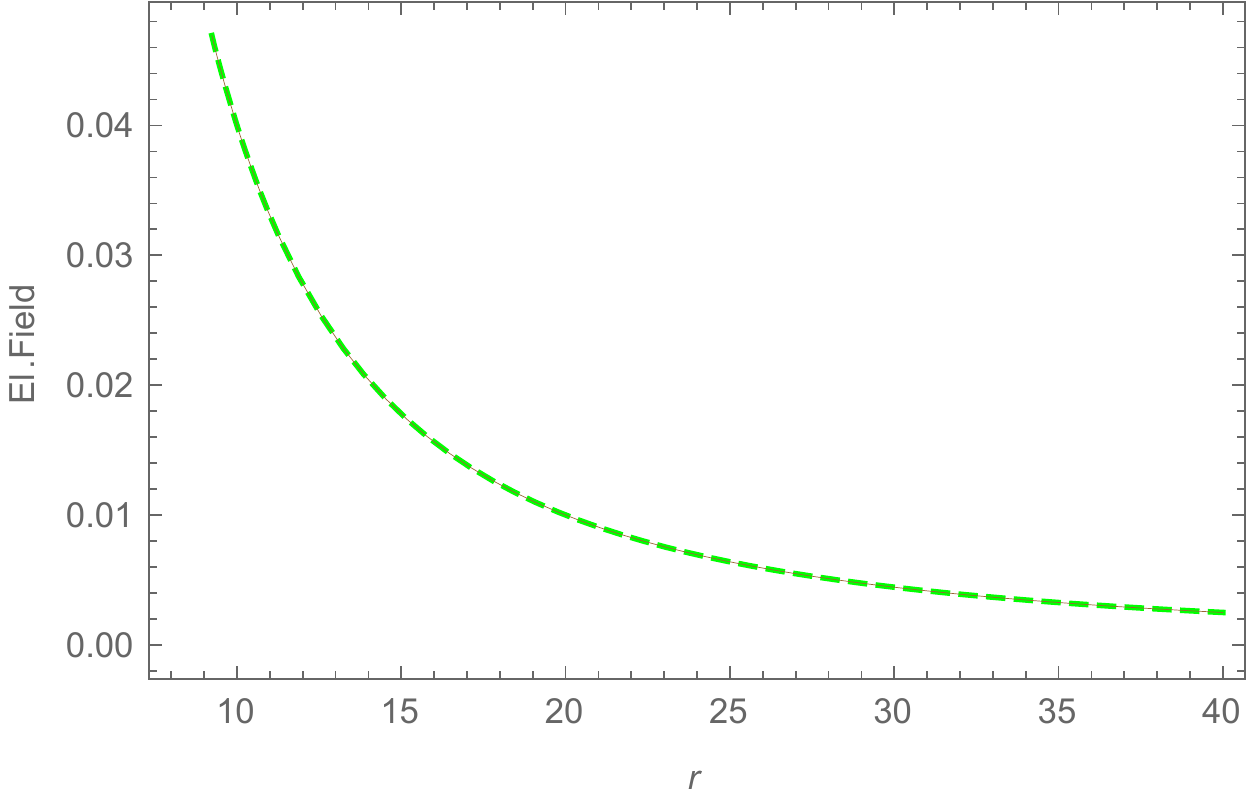}
\includegraphics[width=0.45\columnwidth]{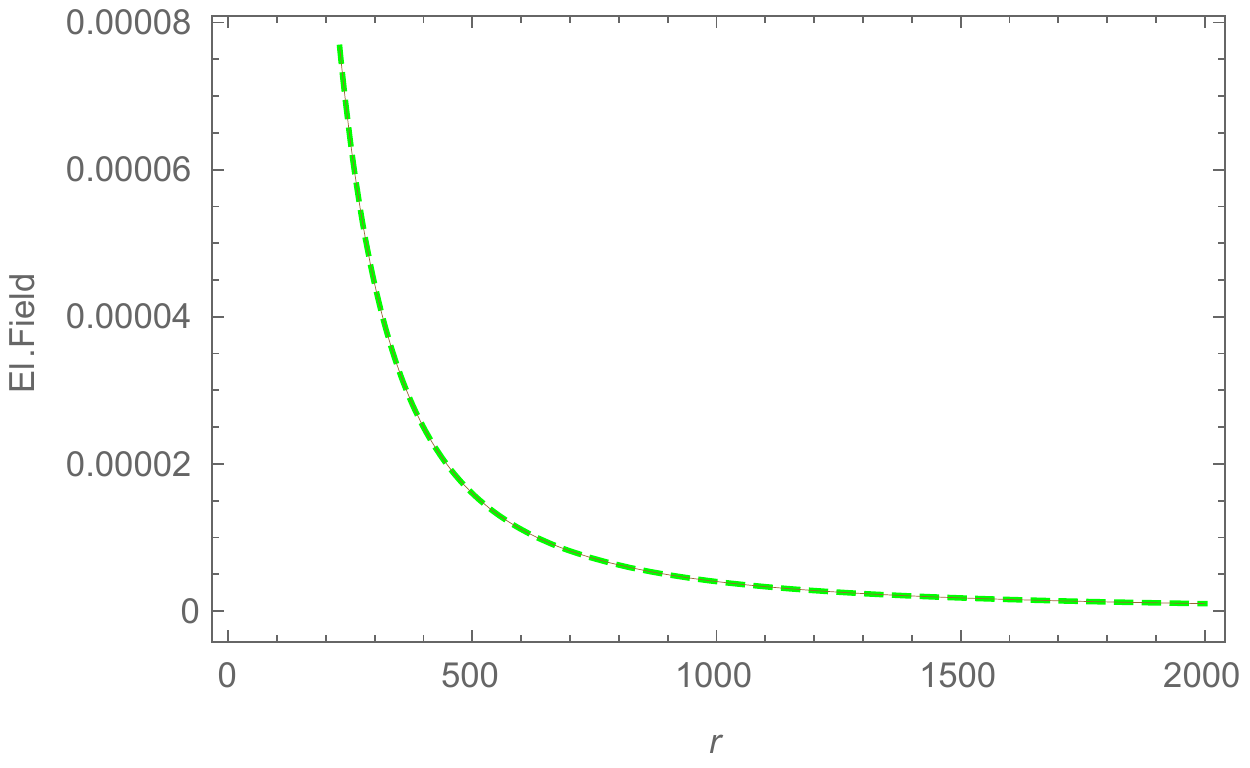}
\caption{Comparative plots of the complete first-order corrected electric field (green) with that of the uncorrected RN electric field (yellow) assuming $M = 5\, , Q = 4$ and $\mu = \frac{m^2}{6} = 10^{-6}$. \newline Left:  From $r=8$ to $r=40$. ~~~~  Right: From  $r=8$ to $r=2000$}
\end{center}
\end{figure}

\subsection{Higher order corrections}
We have just demonstrated that the Kaluza reduced action admits real first-order corrections outside the event horizon of a RN black hole. The complete solution can be derived iteratively and constructed in powers of $\mu = \frac{m^2}{6}$. However, in order for the solutions to be perturbative, the corrections to the metric and the field strength must continue to not grow at higher orders in $\mu$. Here we will consider the nature of the corrections at order $\mu^2$ and argue why this behaviour will continue to hold to all orders. Following the first-order correction, the metric and electromagnetic field strength tensor to order $\mu^2$ can be expressed as
\begin{align}
g_{\mu\nu} &= 
\begin{pmatrix}
- (\tilde{f}(r) + \mu^2 \tilde{A}(r))  & 0 & 0 & 0 \\ 
0 & {(\tilde{f}(r) + \mu^2 (\tilde{A}(r) - \tilde{B}(r)))}^{-1} & 0 & 0\\
0 & 0 & r^2 &  0\\
0 & 0 & 0 & r^2 {\text{sin}}^{2} {\theta}\\
\end{pmatrix}\, , \notag\\
F_{\mu\nu} &= 
\begin{pmatrix}
0 & \tilde{C}(r)+ \mu^2 \tilde{D}(r) & 0 & 0 \\ 
- \tilde{C}(r)-\mu^2 \tilde{D}(r) & 0 & 0 & 0\\
0 & 0 & 0 &  0\\
0 & 0 & 0 & 0\\
\end{pmatrix}\, ,
\label{met.rn2}
\end{align}
where $\tilde{f}(r)$ and $\tilde{C}(r)$ are $g_{00}$ and $F_{01}$ of Eq.~(\ref{sol.mete}) respectively. To determine the second-order corrections of the metric and electromagnetic field strength, we need to solve the first-order constraint equations given by
\begin{align}
\Box^{(0)} \tilde{U}^{(1)} &= - R^{(1)} + F^{(0)}_{\alpha \beta}F^{(1)\alpha \beta}+ F^{(1)}_{\alpha \beta}F^{(0)\alpha \beta} - \Box^{(1)} \tilde{u} \, , \label{sol.ut2} \\
\Box^{(0)} \tilde{S}^{(1)} &= - \tilde{U}^{(1)} - \Box^{(1)} \tilde{s} \, .
\label{sol.us2}
\end{align} 
$\tilde{U}^{(1)}$ and $\tilde{S}^{(1)}$ denote the first-order corrections of $\tilde{U}$ and $\tilde{S}$ respectively. While the Ricci scalar vanishes on the RN background ($R^{(0)} = 0$), its first order correction does not and has the following expression
\begin{equation}
R^{(1)} =  4 \frac{\left(M^2-Q^2 +3 M r\right)}{r^2} +\mathcal{O}(r^{-3}) \, .
\label{kk.hrf}
\end{equation}
The additional $\mathcal{O}(r^{-3})$ terms for $R^{(1)}$ indicate contributions from those terms which were ignored in going from Eq.~(\ref{sol.kkcor}) to Eq.~(\ref{sol.mete}). Denoting the right hand side of Eq.~(\ref{sol.ut2}) by $l(r)$, it is clear that $\Box^{(0)} \tilde{U}^{(1)} = l(r)$ has the same homogeneous solution about the RN background involving the $\text{Tanh}(x)$ function. This piece can always be ignored through the choice of constants. The remaining inhomogeneous solution has the general form
\begin{align}
 \tilde{U}^{(1)}(r) =  a_0 \tilde{u}(r) + \cdots \, ,
\label{kk.conU1}
\end{align}
where $a_0$ is a constant which depends only on the mass and charge of the RN black hole and $\cdots$ represent terms which involve products of logarithms and polylogarithms. These additional contributions are a result of the non-vanishing Ricci scalar at this order and the contribution from $\Box^{(1)} \tilde{u}$. Following the steps outlined in the first-order correction of the previous subsection, we find that when $r_H \ll r$ the leading order corrections at this order are given by
\begin{align}
\tilde{A}(r) &= \alpha r + \beta + \mathcal{O}(r^{-1}) \, , \notag\\
\tilde{B}(r) &= \gamma + \mathcal{O}(r^{-1}) \, ,\notag\\
\tilde{D}(r) &= \frac{\delta}{r} + \mathcal{O}(r^{-2}) \, . 
\label{sol.kkcor2}
\end{align}
In Eq.~(\ref{sol.kkcor2}), $\alpha\,,\beta\,, \gamma$ and $\delta$ are dimensionful constants which depend only on the mass and charge of the RN black hole. Each of the leading contributions indicated in Eq.~(\ref{sol.kkcor2}) are a result of the $a_0 \tilde{u}$ term of $ \tilde{U}^{(1)}(r)$. The product of logarithm terms indicated through the ellipsis in Eq.~(\ref{kk.conU1}) contribute only to the subleading terms in Eq.~(\ref{sol.kkcor2}). It can be noted that the leading contribution in Eq.~(\ref{sol.kkcor2}) has the same form as Eq.~(\ref{sol.kkcor}), up to new constants which involve the parameters of the RN black hole. The dependence on the parameters of black holes imply  $\mu^2 \tilde{A}(r) < \mu A(r)$, $\mu^2 \tilde{B}(r) < \mu B(r)$ and $\mu^2 \tilde{D}(r) < \mu D(r)$, which further imply that the corrections are perturbative up to this order.   \\

Since the solutions for $\tilde{U}(r)\,,\tilde{S}(r)\,, A(r) \,,B(r)$ and $D(r)$ at each order result from the $r$ dependence of the solutions of the previous order, it follows that the corrections at higher orders in $\mu$ will also have the same leading order in $r$ behaviour.
By considering the region outside the horizon and up to subleading terms in $r$, the corrections are perturbative for $r>r_H$ and $\mu r^2 \ll 1$.  
This extends the observation made about the Schwarzschild background, where the nature of the first-order corrections were shown by numeric integration to be maintained at higher orders for the $\mu r^2 \ll 1$ region outside the event horizon of the black hole~\cite{Kehagias:2014sda}. Here we have seen that this property is due to the leading order in $r$ contribution to $\tilde{U}$ at all orders, which up to dimensionful constants, is of the same form as $\tilde{u}$.

\section{Discussion} \label{sec.con}
In this paper, we investigated the coupling of the electromagnetic field to non-local gravity theories and their implications on the resulting electrovacuum solutions. In particular, we considered the action introduced in~\cite{Maggiore:2014sia}, which is causal, covariant and free of ghosts, and satisfies currently known cosmological observations. However, in Sec.~\ref{sec.mm} we also noted that while the action admits real perturbative corrections about the Schwarzschild background, it does not admit similar corrections about the RN background when the electromagnetic field is minimally coupled. This was determined through the constraint equation of the auxilliary field `$U$' given in Eq.~(\ref{mag.aux}), whose introduction was needed for the local formulation of the theory. We determined that this equation does not admit real solutions outside the event horizon of the RN black hole. As a consequence, we cannot construct solutions which involve real, perturbative corrections about the RN background.  Since this result also follows directly from the auxilliary field equation, our conclusion holds for other non-local pure gravitational theories involving the Ricci scalar. One reason for the absence of real solutions of the constraint equation on the RN background is the presence of the electric charge in the metric. We thus require a charged source for the constraint equation, which can only result from non-local terms involving the electromagnetic fields.  

We then considered the modified coupling of the electromagnetic field by performing the Kaluza reduction on the non-local action of Eq.~(\ref{kk.mact1}) in Sec.~(\ref{sec.kk}). The Kaluza reduced action involves terms which are non-local in both gauge and gravitational fields. In Sec.~(\ref{sec.eom}), we introduced auxilliary fields and considered the resulting local formulation of the action. Due to the non-local coupling of the electromagnetic field in the Kaluza reduced action, Maxwell's equation and the stress-energy tensor receive corrections involving the auxilliary fields of the theory.
In order to derive the classical solutions, we provided an iterative approach in Sec.~\ref{sec.sol}, catered to the five coupled differential equations one needs to solve. By considering the RN background, we demonstrated that the auxilliary field now admit real solutions in the region outside the event horizon. This allowed for the derivation of the first-order in $\mu$ corrections of the metric and electromagnetic field strength about the RN background, whose expressions up to leading orders in $r$ was provided in Eq.~(\ref{sol.kkcor}). We then considered higher order corrections to argue that the corrections are perturbative and have the same leading order in $r$ behaviour as the first-order corrections, up to dimensionful constants.   
Thus the Kaluza ansatz provides the non-local coupling of the electromagnetic field needed to admit real perturbative corrections about electrovacuum solutions of GR.

The lowest order solutions of the auxilliary fields, given in Eq.~(\ref{eom.U2}) about the Schwarzschild background for the original non-local theory and Eq.~(\ref{sol.log2}) about the RN background for the Kaluza reduced action, both involved the logarithm of the lapse function of the respective background. This was a consequence of adopting static coordinates, which allowed us to investigate perturbations outside the event horizon of the black hole. To consider the near horizon physics of black holes, it will be interesting to use coordinates which are well behaved across the horizon. This will in particular be relevant to study quantum effects due to non-local fields on curved backgrounds. Here we note the `non-violent non-local' (NVNL) proposal~\cite{Giddings:2012gc,Giddings:2017mym}, which could provide a possible resolution of the information paradox. Further, it has been argued that some of the consequences of this proposal could have observable signatures in future gravitational wave observations~\cite{Giddings:2014ova,Giddings:2016tla,Liebling:2017pqs}. Some implications of the NVNL proposal on non-local scalar fields were considered in \cite{Giddings:2013noa} and it will be interesting to extend these results to non-local gauge fields. As the non-local electromagnetic fields involved in the Kaluza reduced action also modify Maxwell's equations, the Gauss law constraint of the theory will involve non-local corrections as well. This could have further implications on the charges and near horizon properties of black holes.

Non-local electromagnetic fields allow for other quantum effects on curved backgrounds. The anomaly-induced quantum effective actions resulting from background gravitational and gauge fields have applications in the scattering amplitudes on curved backgrounds. Contributions to graviton-photon amplitudes will exist when the effective action contains non-local terms involving an inverse quartic operator as well as both $R$ and $F_{\alpha \beta} F^{\alpha \beta}$~\cite{Giannotti:2008cv,Armillis:2010qk}. Due to the presence of similar terms in the non-local action of Eq.~(\ref{kk.mact2}), one can expect analogous scattering processes to result from the Kaluza reduced action. We look forward to investigating these and related topics in future work. 


\begin{thebibliography}{99}
\bibitem{Perlmutter:1998np} 
  S.~Perlmutter {\it et al.} [Supernova Cosmology Project Collaboration],
  Astrophys.\ J.\  {\bf 517}, 565 (1999)
  doi:10.1086/307221
  [astro-ph/9812133].


\bibitem{Riess:1998cb} 
  A.~G.~Riess {\it et al.} [Supernova Search Team],
  Astron.\ J.\  {\bf 116}, 1009 (1998)
  doi:10.1086/300499
  [astro-ph/9805201].


\bibitem{ArkaniHamed:2002fu} 
  N.~Arkani-Hamed, S.~Dimopoulos, G.~Dvali and G.~Gabadadze,
  hep-th/0209227.

\bibitem{Dvali:2006su} 
  G.~Dvali,
  New J.\ Phys.\  {\bf 8}, 326 (2006)
  doi:10.1088/1367-2630/8/12/326
  [hep-th/0610013].


\bibitem{Barvinsky:2003kg} 
  A.~O.~Barvinsky,
  Phys.\ Lett.\ B {\bf 572}, 109 (2003)
  doi:10.1016/j.physletb.2003.08.055
  [hep-th/0304229].


\bibitem{Dvali:2007kt} 
  G.~Dvali, S.~Hofmann and J.~Khoury,
  Phys.\ Rev.\ D {\bf 76}, 084006 (2007)
  doi:10.1103/PhysRevD.76.084006
  [hep-th/0703027 [HEP-TH]].


\bibitem{Deser:2007jk} 
  S.~Deser and R.~P.~Woodard,
  Phys.\ Rev.\ Lett.\  {\bf 99}, 111301 (2007)
  doi:10.1103/PhysRevLett.99.111301
  [arXiv:0706.2151 [astro-ph]].


\bibitem{Nojiri:2007uq} 
  S.~Nojiri and S.~D.~Odintsov,
  Phys.\ Lett.\ B {\bf 659}, 821 (2008)
  doi:10.1016/j.physletb.2007.12.001
  [arXiv:0708.0924 [hep-th]].


\bibitem{Koivisto:2008xfa} 
  T.~Koivisto,
  Phys.\ Rev.\ D {\bf 77}, 123513 (2008)
  doi:10.1103/PhysRevD.77.123513
  [arXiv:0803.3399 [gr-qc]].


\bibitem{Capozziello:2008gu} 
  S.~Capozziello, E.~Elizalde, S.~Nojiri and S.~D.~Odintsov,
  Phys.\ Lett.\ B {\bf 671}, 193 (2009)
  doi:10.1016/j.physletb.2008.11.060
  [arXiv:0809.1535 [hep-th]].


\bibitem{Deffayet:2009ca} 
  C.~Deffayet and R.~P.~Woodard,
  JCAP {\bf 0908}, 023 (2009)
  doi:10.1088/1475-7516/2009/08/023
  [arXiv:0904.0961 [gr-qc]].




\bibitem{Elizalde:2011su} 
  E.~Elizalde, E.~O.~Pozdeeva and S.~Y.~Vernov,
  Phys.\ Rev.\ D {\bf 85}, 044002 (2012)
  doi:10.1103/PhysRevD.85.044002
  [arXiv:1110.5806 [astro-ph.CO]].



\bibitem{Foffa:2013vma} 
  S.~Foffa, M.~Maggiore and E.~Mitsou,
  Int.\ J.\ Mod.\ Phys.\ A {\bf 29}, 1450116 (2014)
  doi:10.1142/S0217751X14501164
  [arXiv:1311.3435 [hep-th]].


\bibitem{Sandstad:2013oja} 
  M.~Sandstad, T.~S.~Koivisto and D.~F.~Mota,
  Class.\ Quant.\ Grav.\  {\bf 30}, 155005 (2013)
  doi:10.1088/0264-9381/30/15/155005
  [arXiv:1305.0695 [gr-qc]].


\bibitem{Woodard:2014iga} 
  R.~P.~Woodard,
  Found.\ Phys.\  {\bf 44}, 213 (2014)
  doi:10.1007/s10701-014-9780-6
  [arXiv:1401.0254 [astro-ph.CO]].


\bibitem{Barvinsky:2014lja} 
  A.~O.~Barvinsky,
  Mod.\ Phys.\ Lett.\ A {\bf 30}, no. 03n04, 1540003 (2015)
  doi:10.1142/S0217732315400039
  [arXiv:1408.6112 [hep-th]].


\bibitem{Tsamis:2014hra} 
  N.~C.~Tsamis and R.~P.~Woodard,
  JCAP {\bf 1409}, 008 (2014)
  doi:10.1088/1475-7516/2014/09/008
  [arXiv:1405.4470 [astro-ph.CO]].


\bibitem{Soussa:2003vv} 
  M.~E.~Soussa and R.~P.~Woodard,
  Class.\ Quant.\ Grav.\  {\bf 20}, 2737 (2003)
  doi:10.1088/0264-9381/20/13/321
  [astro-ph/0302030].


\bibitem{Hehl:2008eu} 
  F.~W.~Hehl and B.~Mashhoon,
  Phys.\ Lett.\ B {\bf 673}, 279 (2009)
  doi:10.1016/j.physletb.2009.02.033
  [arXiv:0812.1059 [gr-qc]].


\bibitem{Barvinsky:2011hd} 
  A.~O.~Barvinsky,
  Phys.\ Lett.\ B {\bf 710}, 12 (2012)
  doi:10.1016/j.physletb.2012.02.075
  [arXiv:1107.1463 [hep-th]].


\bibitem{Arraut:2013qra} 
  I.~Arraut,
  Int.\ J.\ Mod.\ Phys.\ D {\bf 23}, 1450008 (2014)
  doi:10.1142/S0218271814500084
  [arXiv:1310.0675 [gr-qc]].


\bibitem{Deffayet:2014lba} 
  C.~Deffayet, G.~Esposito-Farese and R.~P.~Woodard,
  Phys.\ Rev.\ D {\bf 90}, no. 6, 064038 (2014)
  Addendum: [Phys.\ Rev.\ D {\bf 90}, no. 8, 089901 (2014)]
  doi:10.1103/PhysRevD.90.089901, 10.1103/PhysRevD.90.064038
  [arXiv:1405.0393 [astro-ph.CO]].


\bibitem{Woodard:2014wia} 
  R.~P.~Woodard,
  Can.\ J.\ Phys.\  {\bf 93}, no. 2, 242 (2015)
  doi:10.1139/cjp-2014-0156
  [arXiv:1403.6763 [astro-ph.CO]].


\bibitem{Wetterich:1997bz} 
  C.~Wetterich,
  Gen.\ Rel.\ Grav.\  {\bf 30}, 159 (1998)
  doi:10.1023/A:1018837319976
  [gr-qc/9704052].


\bibitem{Donoghue:2015nba} 
  J.~F.~Donoghue and B.~K.~El-Menoufi,
  JHEP {\bf 1510}, 044 (2015)
  doi:10.1007/JHEP10(2015)044
  [arXiv:1507.06321 [hep-th]].


\bibitem{Tomboulis:1997gg} 
  E.~T.~Tomboulis,
  hep-th/9702146.


\bibitem{Biswas:2011ar} 
  T.~Biswas, E.~Gerwick, T.~Koivisto and A.~Mazumdar,
  Phys.\ Rev.\ Lett.\  {\bf 108}, 031101 (2012)
  doi:10.1103/PhysRevLett.108.031101
  [arXiv:1110.5249 [gr-qc]].

\bibitem{Modesto:2011kw} 
  L.~Modesto,
  Phys.\ Rev.\ D {\bf 86}, 044005 (2012)
  doi:10.1103/PhysRevD.86.044005
  [arXiv:1107.2403 [hep-th]].


\bibitem{Modesto:2012ys} 
  L.~Modesto,
  Astron. Rev. 8.2 (2013) 4-33
  [arXiv:1202.3151 [hep-th]].


\bibitem{Biswas:2013cha} 
  T.~Biswas, A.~Conroy, A.~S.~Koshelev and A.~Mazumdar,
  Class.\ Quant.\ Grav.\  {\bf 31}, 015022 (2014)
  Erratum: [Class.\ Quant.\ Grav.\  {\bf 31}, 159501 (2014)]
  doi:10.1088/0264-9381/31/1/015022, 10.1088/0264-9381/31/15/159501
  [arXiv:1308.2319 [hep-th]].


\bibitem{Briscese:2013lna} 
  F.~Briscese, L.~Modesto and S.~Tsujikawa,
  Phys.\ Rev.\ D {\bf 89}, no. 2, 024029 (2014)
  doi:10.1103/PhysRevD.89.024029
  [arXiv:1308.1413 [hep-th]].


\bibitem{Modesto:2014lga} 
  L.~Modesto and L.~Rachwal,
  Nucl.\ Phys.\ B {\bf 889}, 228 (2014)
  doi:10.1016/j.nuclphysb.2014.10.015
  [arXiv:1407.8036 [hep-th]].


\bibitem{Talaganis:2014ida} 
  S.~Talaganis, T.~Biswas and A.~Mazumdar,
  Class.\ Quant.\ Grav.\  {\bf 32}, no. 21, 215017 (2015)
  doi:10.1088/0264-9381/32/21/215017
  [arXiv:1412.3467 [hep-th]].


\bibitem{Tomboulis:2015esa} 
  E.~T.~Tomboulis,
  Mod.\ Phys.\ Lett.\ A {\bf 30}, no. 03n04, 1540005 (2015).
  doi:10.1142/S0217732315400052


\bibitem{Modesto:2017sdr} 
  L.~Modesto and L.~Rachwał,
  Int.\ J.\ Mod.\ Phys.\ D {\bf 26}, no. 11, 1730020 (2017).
  doi:10.1142/S0218271817300208


\bibitem{Giddings:2009ae} 
  S.~B.~Giddings,
  Class.\ Quant.\ Grav.\  {\bf 28}, 025002 (2011)
  doi:10.1088/0264-9381/28/2/025002
  [arXiv:0911.3395 [hep-th]].


\bibitem{Giddings:2012gc} 
  S.~B.~Giddings,
  Phys.\ Rev.\ D {\bf 88}, 064023 (2013)
  doi:10.1103/PhysRevD.88.064023
  [arXiv:1211.7070 [hep-th]].


\bibitem{Giddings:2017mym} 
  S.~B.~Giddings,
  JHEP {\bf 1712}, 047 (2017)
  doi:10.1007/JHEP12(2017)047
  [arXiv:1701.08765 [hep-th]].


\bibitem{Liebling:2017pqs} 
  S.~L.~Liebling, M.~Kavic and M.~Lippert,
  arXiv:1707.02299 [gr-qc].


\bibitem{Buchbinder:1992rb} 
  I.~L.~Buchbinder, S.~D.~Odintsov and I.~L.~Shapiro,
  Bristol, UK: IOP (1992) 413 p


\bibitem{Shapiro:2008sf} 
  I.~L.~Shapiro,
  Class.\ Quant.\ Grav.\  {\bf 25}, 103001 (2008)
  doi:10.1088/0264-9381/25/10/103001
  [arXiv:0801.0216 [gr-qc]].


\bibitem{Deser:2013uya} 
  S.~Deser and R.~P.~Woodard,
  JCAP {\bf 1311}, 036 (2013)
  doi:10.1088/1475-7516/2013/11/036
  [arXiv:1307.6639 [astro-ph.CO]].

\bibitem{Maggiore:2016gpx} 
  M.~Maggiore,
  Fundam.\ Theor.\ Phys.\  {\bf 187}, 221 (2017)
  doi:10.1007/978-3-319-51700-116
  [arXiv:1606.08784 [hep-th]].

\bibitem{Maggiore:2014sia} 
  M.~Maggiore and M.~Mancarella,
  Phys.\ Rev.\ D {\bf 90}, no. 2, 023005 (2014)
  doi:10.1103/PhysRevD.90.023005
  [arXiv:1402.0448 [hep-th]].

\bibitem{Nesseris:2014mea} 
  S.~Nesseris and S.~Tsujikawa,
  Phys.\ Rev.\ D {\bf 90}, no. 2, 024070 (2014)
  doi:10.1103/PhysRevD.90.024070
  [arXiv:1402.4613 [astro-ph.CO]].


\bibitem{Barreira:2014kra} 
  A.~Barreira, B.~Li, W.~A.~Hellwing, C.~M.~Baugh and S.~Pascoli,
  JCAP {\bf 1409}, no. 09, 031 (2014)
  doi:10.1088/1475-7516/2014/09/031
  [arXiv:1408.1084 [astro-ph.CO]].


\bibitem{Dirian:2014xoa} 
  Y.~Dirian and E.~Mitsou,
  JCAP {\bf 1410}, no. 10, 065 (2014)
  doi:10.1088/1475-7516/2014/10/065
  [arXiv:1408.5058 [gr-qc]].


\bibitem{Dirian:2014ara} 
  Y.~Dirian, S.~Foffa, N.~Khosravi, M.~Kunz and M.~Maggiore,
  JCAP {\bf 1406}, 033 (2014)
  doi:10.1088/1475-7516/2014/06/033
  [arXiv:1403.6068 [astro-ph.CO]].


\bibitem{Dirian:2014bma} 
  Y.~Dirian, S.~Foffa, M.~Kunz, M.~Maggiore and V.~Pettorino,
  JCAP {\bf 1504}, no. 04, 044 (2015)
  doi:10.1088/1475-7516/2015/04/044
  [arXiv:1411.7692 [astro-ph.CO]].


\bibitem{Dirian:2016puz} 
  Y.~Dirian, S.~Foffa, M.~Kunz, M.~Maggiore and V.~Pettorino,
  JCAP {\bf 1605}, no. 05, 068 (2016)
  doi:10.1088/1475-7516/2016/05/068
  [arXiv:1602.03558 [astro-ph.CO]].


\bibitem{Nersisyan:2016hjh} 
  H.~Nersisyan, Y.~Akrami, L.~Amendola, T.~S.~Koivisto and J.~Rubio,
  Phys.\ Rev.\ D {\bf 94}, no. 4, 043531 (2016)
  doi:10.1103/PhysRevD.94.043531
  [arXiv:1606.04349 [gr-qc]].


\bibitem{Belgacem:2017cqo} 
  E.~Belgacem, Y.~Dirian, S.~Foffa and M.~Maggiore,
  arXiv:1712.07066 [hep-th].



\bibitem{Nersisyan:2016jta} 
  H.~Nersisyan, Y.~Akrami, L.~Amendola, T.~S.~Koivisto, J.~Rubio and A.~R.~Solomon,
  Phys.\ Rev.\ D {\bf 95}, no. 4, 043539 (2017)
  doi:10.1103/PhysRevD.95.043539
  [arXiv:1610.01799 [gr-qc]].


\bibitem{Fernandes:2014bka} 
  K.~Fernandes and A.~Lahiri,
  Phys.\ Rev.\ D {\bf 91}, no. 4, 044014 (2015)
  doi:10.1103/PhysRevD.91.044014
  [arXiv:1405.2172 [gr-qc]].


\bibitem{Bamba:2013fta} 
  K.~Bamba, S.~Nojiri and S.~D.~Odintsov,
  Phys.\ Lett.\ B {\bf 725}, 368 (2013)
  doi:10.1016/j.physletb.2013.07.052
  [arXiv:1304.6191 [gr-qc]].

\bibitem{Ashoorioon:2004rs} 
  A.~Ashoorioon and R.~B.~Mann,
  Phys.\ Rev.\ D {\bf 71}, 103509 (2005)
  doi:10.1103/PhysRevD.71.103509
  [gr-qc/0410053].


\bibitem{Golovnev:2008cf} 
  A.~Golovnev, V.~Mukhanov and V.~Vanchurin,
  JCAP {\bf 0806}, 009 (2008)
  doi:10.1088/1475-7516/2008/06/009
  [arXiv:0802.2068 [astro-ph]].


\bibitem{Bamba:2008ja} 
  K.~Bamba and S.~D.~Odintsov,
  JCAP {\bf 0804}, 024 (2008)
  doi:10.1088/1475-7516/2008/04/024
  [arXiv:0801.0954 [astro-ph]].


\bibitem{Sadeghi:2009pu} 
  J.~Sadeghi, M.~R.~Setare and A.~Banijamali,
  Eur.\ Phys.\ J.\ C {\bf 64}, 433 (2009)
  doi:10.1140/epjc/s10052-009-1152-6
  [arXiv:0906.0713 [hep-th]].


\bibitem{Harko:2014gwa} 
  T.~Harko and F.~S.~N.~Lobo,
  Galaxies {\bf 2}, no. 3, 410 (2014)
  doi:10.3390/galaxies2030410
  [arXiv:1407.2013 [gr-qc]].


\bibitem{Drummond:1979pp} 
  I.~T.~Drummond and S.~J.~Hathrell,
  Phys.\ Rev.\ D {\bf 22}, 343 (1980).
  doi:10.1103/PhysRevD.22.343


\bibitem{Bastianelli:2004zp} 
  F.~Bastianelli and C.~Schubert,
  JHEP {\bf 0502}, 069 (2005)
  doi:10.1088/1126-6708/2005/02/069
  [gr-qc/0412095].


\bibitem{Bastianelli:2007jv} 
  F.~Bastianelli, U.~Nucamendi, C.~Schubert and V.~M.~Villanueva,
  JHEP {\bf 0711}, 099 (2007)
  doi:10.1088/1126-6708/2007/11/099
  [arXiv:0710.5572 [gr-qc]].


\bibitem{Bastianelli:2012bz} 
  F.~Bastianelli, O.~Corradini, J.~M.~Dávila and C.~Schubert,
  Phys.\ Lett.\ B {\bf 716}, 345 (2012)
  doi:10.1016/j.physletb.2012.08.030
  [arXiv:1202.4502 [hep-th]].


\bibitem{Kehagias:2014sda} 
  A.~Kehagias and M.~Maggiore,
  JHEP {\bf 1408}, 029 (2014)
  doi:10.1007/JHEP08(2014)029
  [arXiv:1401.8289 [hep-th]].

\bibitem{Giddings:2014ova} 
  S.~B.~Giddings,
  Phys.\ Rev.\ D {\bf 90}, no. 12, 124033 (2014)
  doi:10.1103/PhysRevD.90.124033
  [arXiv:1406.7001 [hep-th]].


\bibitem{Giddings:2016tla} 
  S.~B.~Giddings,
  Class.\ Quant.\ Grav.\  {\bf 33}, no. 23, 235010 (2016)
  doi:10.1088/0264-9381/33/23/235010
  [arXiv:1602.03622 [gr-qc]].


\bibitem{Giddings:2013noa} 
  S.~B.~Giddings and Y.~Shi,
  Phys.\ Rev.\ D {\bf 89}, no. 12, 124032 (2014)
  doi:10.1103/PhysRevD.89.124032
  [arXiv:1310.5700 [hep-th]].


\bibitem{Giannotti:2008cv} 
  M.~Giannotti and E.~Mottola,
  Phys.\ Rev.\ D {\bf 79}, 045014 (2009)
  doi:10.1103/PhysRevD.79.045014
  [arXiv:0812.0351 [hep-th]].


\bibitem{Armillis:2010qk} 
  R.~Armillis, C.~Coriano and L.~Delle Rose,
  Phys.\ Rev.\ D {\bf 82}, 064023 (2010)
  doi:10.1103/PhysRevD.82.064023
  [arXiv:1005.4173 [hep-ph]].
\end{thebibliography}
\end{document}